\documentclass[preprint,3p,twocolumn]{elsarticle}

\usepackage{graphics}
\usepackage{epstopdf}
\usepackage{epsfig}
\usepackage{widetext}
\usepackage{flushend}
\usepackage{cuted}

\usepackage{amssymb}
\usepackage{color}

\journal{JQSRT}

\begin{document}

\begin{frontmatter}



\title{Robustness of the Fractal Regime for the Multiple-Scattering Structure Factor}


\author{Nisha Katyal}

\address{School of Physical Sciences, Jawaharlal Nehru University, New Delhi 110067, India}
\author{ Robert Botet}

\address{Laboratoire de Physique des Solides, CNRS UMR8502, Universit\'e Paris-Sud, Universit\'e Paris-Saclay, 91405 Orsay, France}
\author{Sanjay Puri}

\address{School of Physical Sciences, Jawaharlal Nehru University, New Delhi 110067, India}

\begin{abstract}

In the single-scattering theory of electromagnetic radiation, the {\it fractal regime} is a definite range in the photon momentum-transfer $q$, which is characterized by the scaling-law behavior of the structure factor: $S(q) \propto 1/q^{d_f}$. This allows a straightforward estimation of the fractal dimension $d_f$ of aggregates in {\it Small-Angle X-ray Scattering} (SAXS) experiments. However, this behavior is not commonly studied in optical scattering experiments because of the lack of information on its domain of validity. In the present work, we propose a definition of the multiple-scattering structure factor, which naturally generalizes the single-scattering function $S(q)$. We show that the mean-field theory of electromagnetic scattering provides an explicit condition to interpret the significance of multiple scattering. In this paper, we investigate and discuss electromagnetic scattering by three classes of fractal aggregates. The results obtained from the TMatrix method show that the fractal scaling range is divided into two domains: 1) a genuine fractal regime, which is robust; 2) a possible anomalous scaling regime, $S(q) \propto 1/q^{\delta}$, with exponent $\delta$ independent of $d_f$, and related to the way the scattering mechanism uses the local morphology of the scatterer. The recognition, and an analysis, of the latter domain is of importance because it may result in significant reduction of the fractal regime, and brings into question the proper mechanism in the build-up of multiple-scattering.

\end{abstract}


\end{frontmatter}


\section{Introduction}
\label{s1}

The investigation of the optical response of disordered matter is an intricate problem. The most common tools to extract information from the scattering data are the trial-and-error methods. However, due to a large number of relevant parameters, the results are tedious to obtain and may prove to be ambiguous. For this reason, direct methods allowing computation of system parameters from the scattering data are highly desirable. 

In the single-scattering case, a few direct methods are known. For example, the fractal dimension of a finite aggregate of particles is given by a special scaling law of the {\it structure factor}, $S(q)$, in the fractal regime \cite{teixeira}:
\begin{equation}
S(\it q) \propto (qa)^{-d_f}.
\label{0}
\end{equation}
Here, $q = |\bf q|$  is the magnitude of the momentum transfer, $a$ is the typical microscopic particle size, and $d_f$ is the fractal dimension of the aggregate. This fractal scaling law is valid in the range $1/R_g < q < 1/a$, where $R_g$ is the mean radius of gyration. The above relation in $q$-space reflects the corresponding scaling law of the pair-correlation function, $g(r)$, in  real space \cite{teixeira}:
\begin{equation}
 g(r) \propto r^{\,d_f-3},
\end{equation}
valid in the range $ a < r < R_g$.

From the introduction of X-ray generators a century ago \cite{xrays} to the present synchrotron facilities \cite{synchro}, the structure factor has turned out to be a major tool to characterize correlations between the particle positions over many length scales \cite{struc}. The relation (\ref{0}) is widely used in small-angle X-ray scattering (SAXS) \cite{xrays} to study the morphology of fractal aggregates \cite{sorensen} or fractal surfaces \cite{saxs}. We can also obtain the specific surface of the system from the {\it Porod regime} \cite{porod}, and the mean gyration radius from the  {\it Guinier regime} \cite{guinier}.  

On the other hand, most light-scattering experiments involve multiple-scattering processes, which are intractable using simple mathematical tools. However, previous studies have shown that Eq.~(\ref{0}) may remain valid for scattering measurements even though multiple-scattering events are suspected to be present \cite{matuda,chen}. Experimentalists widely use the above single-scattering scaling law for the characterization of various fractal aggregates \cite{cai}, though they also claim that minor differences exist between the experimental data and the transmission electron microscopy (TEM) analysis of the fractal dimension. Other results suggest that Eq.~(\ref{0}) might not be valid when multiple-scattering is present. A theoretical argument due to Berry and Percival \cite{berry} claims that the scaling relation in Eq.~(\ref{0}) should fail for aggregates with mass-fractal dimension $d_f > 2$. However, it is not clear what relation should replace Eq.~(\ref{0}) in that case. Results of numerical computations \cite{oh.soren,nisha} suggest that Eq.~(\ref{0}) is valid for the case of $d_f > 2$, but with an effective exponent different from the fractal dimension. 

In this paper, we present detailed theoretical and numerical results to clarify this important issue. The paper is organized as follows. In Section~\ref{s2}, we introduce the static structure factor with multiple-scattering condition. Section~\ref{s3} describes the mean-field multiple-scattering theory for the dipolar and multipolar regimes. In Section~\ref{s4}, we compare theoretical and numerical results for three different types of cluster aggregates. In Section~\ref{sec5}, we discuss the scaling behavior of the structure factor. Finally, we conclude with a summary in Section~\ref{s6}.

\section{General Definition of the Static Structure Factor}
\label{s2}

Consider a monochromatic electromagnetic wave of amplitude $E_0$ and wavelength $\lambda$, illuminating an aggregate of $N$ spherical particles of radius $a$. We assume elastic light-scattering, with the wave-vector ${\bf k}_{\rm inc}$ (of modulus $k=2\pi/\lambda$) defining the propagation direction of the incident beam. The scattered wave-vector, ${\bf k}_{\rm sca}$ (with the same modulus $k$), defines the observation direction. For the general case of a randomly-oriented aggregate, the intensity $I_N({\bf q})$ of the scattered wave depends on $qa$, where $q$ is the magnitude  of the scattering vector ${\bf q} = {\bf k}_{\rm inc} - {\bf k}_{\rm sca}$.

Assuming separation of the optical properties and the spatial distribution of the particles,  $I_N(\bf q)$ is related to the structure factor $S_N({\bf q})$ as
\begin{equation}
I_N({\bf q}) = |E_0|^2 f(N) F({\bf q}) S_N({\bf q})~,
\label{SF1}
\end{equation}
the form factor $F({\bf q})$ being the intensity scattered by a single particle. 

We make the following observations: \\
1) The structure factor $S_N({\bf q})$ is a positive function  which can be conveniently normalized such that $S_N({\bf 0}) = 1$. This function contains information about the spatial distribution of the particles. For the case of a single particle ({\it i.e.}, $N=1$), $S_1({\bf q}) = 1$. \\
2) The scaling factor $f(N)$ is related to the forward-scattering scaling as 
\begin{equation}
f(N) = \frac{I_N({\bf 0})}{|E_0|^2 F({\bf 0})}. 
\end{equation}
In most cases, this function behaves as a power-law: $f(N) \propto N^{\alpha}$, with an exponent $0 < \alpha \le 2$ \cite{sorensen}.

The simplest way to define the structure factor is then via the normalized ratio of the scattered intensity from the $N$ particles ($I_N$) and the scattered intensity from a single particle ($I_1$):
\begin{equation}
S_N({\bf q}) = \frac{I_N({\bf q})}{I_N({\bf 0})} \cdot \frac{I_1({\bf 0})}{I_1({\bf q})} .
\label{SF2}
\end{equation}
Now, $I_N$ being written as the product of a function of the optical parameters and a function involving the spatial distribution of the particles, the definition (\ref{SF2}) results in a quantity which depends essentially on the mass distribution of the aggregate. \\

The Rayleigh-Debye-Gans (RDG) theory provides a framework in which the separation between the optical and geometrical properties is realized \cite{rgd}. Indeed, the RDG theory tells us that the single-scattering of the incoming wave by a collection of $N$ electromagnetic dipoles is the dominant process due to the weak electric polarizability of the particles inside the aggregate. The structure factor in Eq.~(\ref{SF2}) is then written as the square modulus of the Fourier transform of the density distribution of the scattering system \cite{struc}:
\begin{eqnarray}
S_N({\bf q}) & = & \left| \frac{1}{N} \sum_{j=1}^{N} e^{i {\bf q} \cdot {\bf r}_j} \right|^2 \label{2} \\
& = & \frac{1}{N} \left( 1 + \rho \int (g({\bf r}) - 1) e^{i {\bf q} \cdot {\bf r}} d{\bf r} \right) \label{Sg} .
\end{eqnarray}
In Eq.~(\ref{2}), {$\bf r_j$} denotes the position of the $j^{th}$ particle in the aggregate. In Eq.~(\ref{Sg}), $g({\bf r})$ is the pair-correlation function \cite{teixeira}, and $\rho = N/V$ is the particle number density in a given volume $V$.\\
 
For a fractal scatterer of radius of gyration $R_g$, which is an aggregate of spherical particles of  radius $a$, the main features of the RDG structure factor are as follows\\
1) $qa < a/R_g$ is the  {\it Guinier regime} \cite{guinier}, which depends only on the parameter $qR_g$.\\
2) $a/R_g < qa < 1$ is the {\it fractal regime}, which is characterized by Eq.~(\ref{0}). We focus on this regime in the next section.\\
3) $1 \ll qa$ is the  {\it Porod regime} \cite{porod}, which is not relevant for the present paper.

\section{Mean-Field Approach to the Multiple-Scattering Process}
\label{s3}

For experimental studies, it is important to bridge the gap between the simple single-scattering approximation and the complicated multiple-scattering computations. In this context, the {\it  mean-field} multiple-scattering approximation provides a mathematically tractable method, which includes the main features of the scattering problem. In this approximation, all the particles of the aggregate are expected to radiate the same electromagnetic field, differing only in the phase \cite{berry}. However, the mean-field approach is not consistent with localization phenomena such as coherent scattering \cite{muino}.

\subsection{The Dipolar Theory}
\label{s31}

Let us consider an aggregate made up of particles of radius $a$ and complex refractive index $n$. The optical size parameter of the aggregate is $x = ka = 2\pi a/\lambda$. Then, the dipolar regime corresponds to \cite{zubko}
\begin{equation}
|n|x <1 \label{nx}.
\end{equation}
In this regime, the multiple-scattering field is similar to the single-scattering field with the renormalized amplitude \cite{berry}
\begin{equation}
E_0 \rightarrow  \frac{E_0}{1- (N-1) \chi(x) A(k R_g)} \label{renorm} ~.
\end{equation}
Here, $\chi (x)$ is the material constant given by
\begin{equation}
\chi(x)=\frac{3i}{4} \left\{ 1- \exp\left[ \frac{4i}{3} \cdot \left(\frac{n^2-1}{n^2+2}\right) x^3 \right] \right\} \label{chi}.
\end{equation}
Further, the complex-valued coefficient $A$ is a known function of $kR_g$ (not of $qa$), given by Eq.~(\ref{A}) \cite{berry}.

\begin{widetext}
\begin{eqnarray}[!ht]
A(kR_g) &=& \frac{2}{N(N-1)} \sum_{j=1}^{N-1} \sum_{l=j+1}^{N} \frac{e^{i k r_{jl}}}{k r_{jl}} \left[\frac{\sin (kr_{jl})}{kr_{jl}}\left( 1+ \frac{i}{kr_{jl}} - \frac{1}{(kr_{jl})^2} \right) \right. \nonumber \\
& & - \left. \frac{\sin (kr_{jl}) - kr_{jl} \cos (kr_{jl})}{(kr_{jl})^3} \left( 1+\frac{3i}{kr_{jl}}-\frac{3}{(kr_{jl})^2} \right) \right] \label{A} .
\end{eqnarray}
\end{widetext}

From Eq.~(\ref{renorm}), the precise condition for the single-scattering approximation to be valid is written as
\begin{equation}
(N-1) |\chi(x) A(k R_g)| \ll 1 . \label{cond1}
\end{equation} 
Otherwise, the renormalized coefficient has to be accounted for in the scattering formulae. As the coefficient $(N-1) \chi A$ in Eq.~(\ref{renorm}) does not depend upon ${\bf q}$, we deduce an important consequence, i.e., Eq.~(\ref{Sg}) remains valid in the mean-field {\it dipolar} multiple-scattering approximation.

\subsection{The Multipolar Theory}
\label{s32}
 
The mean-field scattering theory in the multipolar case \cite{MFBotet} is comparatively more complicated than the dipolar case. It gives two different factors to renormalize the coefficients in the expansion of the electromagnetic field in terms of the spherical vector wave functions. Consequently, it is natural to question  the validity of the simple Eq.~(\ref{Sg}) in this case. 

Actually, the structure factor can be written in this case as (see Eq.~(25) of Ref.~\cite{MFBotet})
\begin{equation}
S_N({\bf q}) = \frac{I_1^\star ({\bf q})}{I_1 ({\bf q})} \cdot \frac{I_1 ({\bf 0})}{I_1^\star ({\bf 0})} \left| \frac{1}{N} \sum_{j=1}^{N} e^{i {\bf q} \cdot {\bf r}_j} \right|^2 \label{2*} ~,
\end{equation}
where $I_1$ is the intensity of the wave scattered by an isolated sphere, and $I_1^\star$ is the corresponding intensity with the renormalized multipolar coefficients. Then, $S_N({\bf q})$ consists of the RDG structure factor (\ref{2}) corrected by a (possibly complicated) function of $\bf q$. Thus, the fractal scaling relation (\ref{0}) is not guaranteed to work in this framework. \\

It is shown in Ref.~\cite{berry} that the left-hand term of the condition in Eq.~(\ref{cond1}) tends to a constant  $\simeq (n^2-1)(ka)^{3-d_f}$ (independent of the aggregate size) when $d_f < 2$, i.e., fractal dimensions corresponding to very fluffy aggregates. If this constant is small enough, single-scattering is the dominant process regardless of the cluster size. On the other hand, when $d_f > 2$ corresponding to dense aggregates, the left-hand term $\simeq (ka) N^{1-2/d_f}$. Thus, there is a typical cluster size beyond which multiple-scattering cannot be neglected for $d_f > 2$.

\section{Manifestation of Scaling Laws in the Structure Factors of DLCA and DLA Clusters}
\label{s4}

In this section, we study the structure factors of two classes of fractal aggregates of spheres, viz., {\it Diffusion-Limited Cluster Aggregates} or DLCA \cite{me} with fractal dimension $d_f=1.78$; and {\it Diffusion-Limited Aggregates} or DLA \cite{meakin} with $d_f=2.5$. For numerical computation, we use the multiple scattering code TMatrix \cite{mm1990,mm2011}, which takes into account the multipolar terms of the scattered wave.

\subsection{The DLCA Model: An Example with $d_f < 2$}\label{s41}

\begin{figure}[htb]
\centering
\includegraphics[ width=0.4\linewidth]{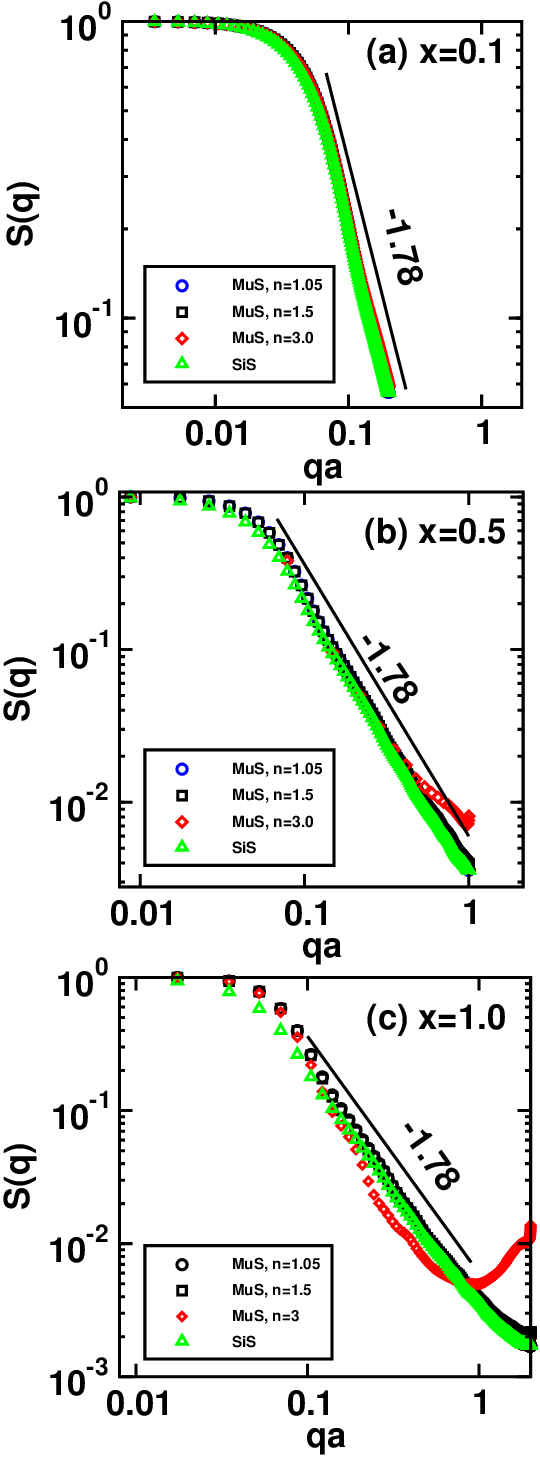}
\caption{Structure factors obtained from multiple-scattering theory or MuS using the TMatrix code for DLCA clusters. These are compared with the single-scattering result or SiS in Eq.~(\ref{2}), denoted by a solid line. Three sets of the optical size parameter are used: (a) $x=0.1$; (b) $x=0.5$; and (c) $x=1.0$. In these log-log plots, a line of slope $-d_f$ (with $d_f=1.78$) is shown for comparison in the range $a/R_g < q a < 1$.}
\label{fig1}
\end{figure}

Fig.~\ref{fig1}(a)-(c) shows TMatrix scattering data [$S(q)$ vs. $q$ on a log-log plot] obtained for the DLCA model. The data is angularly averaged over 20 independent random off-lattice DLCA aggregates of $N=384$ spherical particles. We considered three different values of the optical size parameter: (a) $x=0.1$; (b) $x=0.5$; (c) $x=1.0$, corresponding to wavelengths respectively larger than, of the same order, and smaller than the aggregate gyration radius $R_g$. Moreover, three different values of the refractive index ($n=1.05, 1.5, 3.0$) were used.

\begin{figure}[htb]
\centering
\vspace{0.4cm}
\includegraphics[ width=1\linewidth]{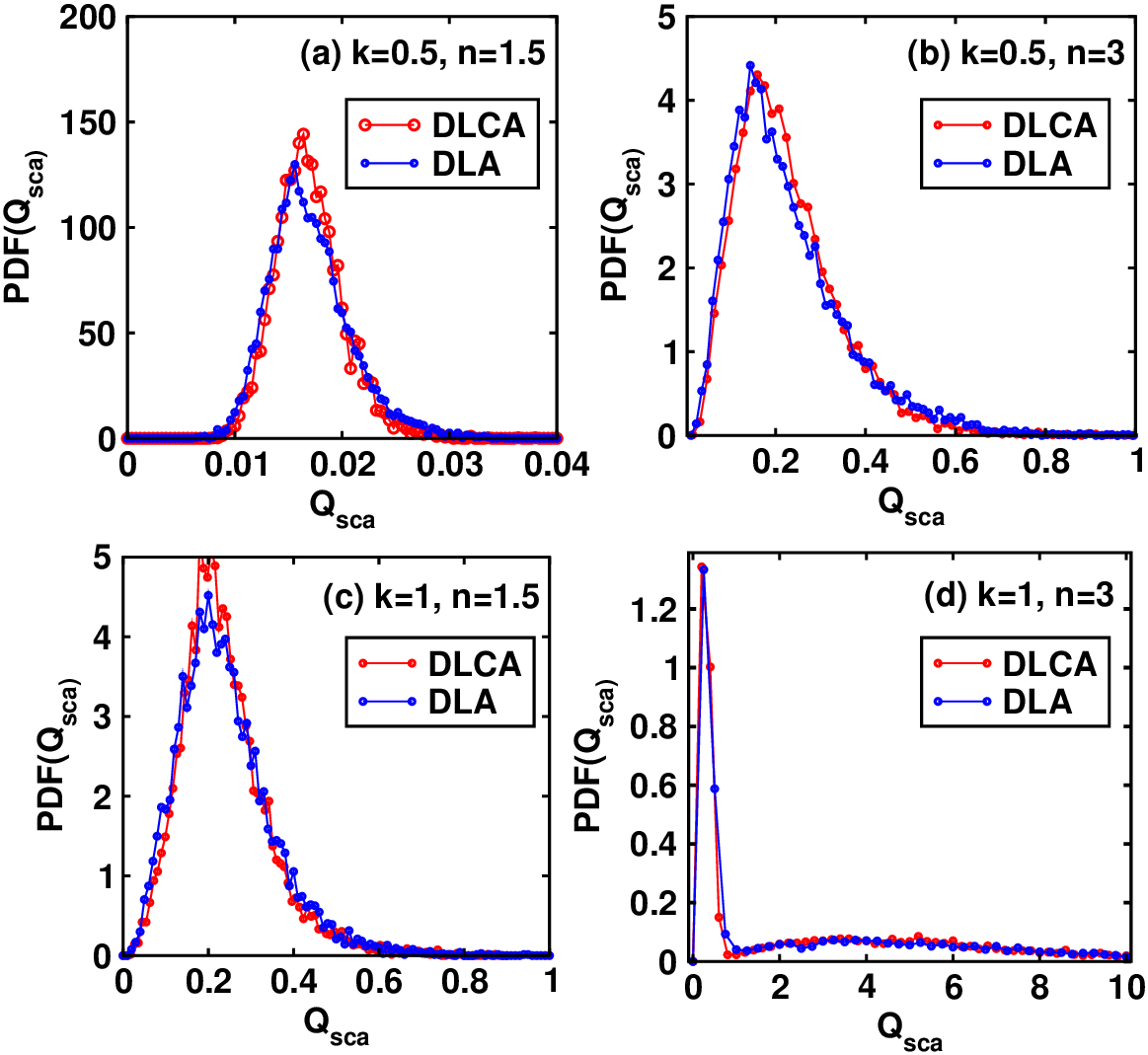}
\caption{Probability distribution functions $P(Q_{\rm sca})$ of the individual scattered efficiencies $Q_{\rm sca}(j)$ of the spheres inside DLA and DLCA aggregates of $N=384$ particles. Data are shown for two different values of the optical parameter $x=ka$ and of the real refractive index $n$: (a) $x=0.5,n=1.5$; (b) $x=0.5,n=3.0$; (c) $x=1.0,n=1.5$; and (d) $x=1.0,n=3.0$. The widths of $P(Q_{\rm sca})$ are significantly smaller than their modes for all the cases except for $(x=1.0~,~n=3.0)$. The values of the mean and the mode ($\left\langle Q_{\rm sca} \right\rangle, Q_m$) are as follows: (a) (0.02~,~0.02); (b) (0.22~,~0.16); (c) (0.23~,~0.19); (d) (2.8~,~0.2) for the most probable mode, and (2.8~,~3.4) for the secondary mode. The distributions in (a)-(c) are clearly unimodal.}
\label{fig2}
\end{figure}

One can check the validity of the mean-field approximation by direct comparison with the exact TMatrix results. For this, we analyze the probability distributions of the particle scattering cross-sections $Q_{\rm sca}(j)$, for all the particles in an aggregate (the index $j$ representing the sphere label). These distributions are shown in Fig.~\ref{fig2} for both DLCAs and DLAs- the latter will be discussed in Sec.~\ref{s42}. The main conclusions are as follows. In all the  cases, except for the largest optical size parameter and largest refractive index $(x=1.0,~n=3.0)$, the mean values $\left\langle Q_{\rm sca} \right\rangle$ agree well with the dominant modes $Q_m$. Thus, the mean-field approach is valid in those cases. This is also consistent with the small values of the quantity  $(N-1) |\chi A|$ (see Table \ref{table1}), which arises in Eq.~(\ref{cond1}). For the case $(x=1.0,~n=3.0)$, the mean-field assumption is not valid, as is clear from  the wide distribution of the individual $Q_{\rm sca}$, and the relatively high value ($\simeq 0.8$) of $(N-1) |\chi A|$. 

\begin{table}
\begin{center}
\begin{tabular}{|c|c|c|c|}
\hline
$x$ \textbackslash ~ $n$ & 1.05 & 1.5 & 3.0 \\ \hline
0.1 &\bf{0.002} & \bf{0.022} & \bf{0.054} \\ \hline
0.5 & \bf{0.029} & \bf{0.174} & 0.431 \\ \hline
1.0 & \bf{0.039} & 0.342 & 0.817 \\ \hline
\end{tabular}
\end{center}
\caption{Values of $(N-1) |\chi A|$ [see Eq.~(\ref{cond1})] for DLCA clusters of $N=384$ particles. We consider various combinations of $x$ and $n$. The dipolar single-scattering theory applies for the values given in boldface.}
\label{table1}
\end{table}

In Fig.\ref{fig1}(a)-(c), we recover the fractal scaling law (\ref{0}) in all cases except for $(x=1.0,~n=3.0)$, on the expected range of values of $qa$, viz., $0.04 < qa < 1$ (for $R_g \simeq 25.4 a$, here). As the single-scattering process is dominant here, the results are in accordance with \cite{chen,frey}.

On the other hand, the failure of the fractal scaling law for $(x=1.0,~n=3.0)$ can be explained by both the high value of $(N-1) |\chi A|$ (relevance of multiple scattering) and the high value of $|n|x$ (multipolar theory applies). We shall discuss this case in greater detail in Sec.~\ref{s42} and Sec.~\ref{sec5}.

\subsection{The DLA Model: An Example with $d_f > 2$}\label{s42}

Next, we consider the DLA model for which $d_f \simeq 2.5$. The computations described in Sec.~\ref{s41} have been repeated for random off-lattice DLA aggregates (with $N=384$ particles). We averaged $S({\bf q})$ over 20 independent realizations, and studied the same values of $x$ and $n$ as in Sec.~\ref{s41}.


Table~\ref{table2} shows the values of $(N-1) |\chi A|$ for the DLA clusters. The distributions of the individual scattering cross-sections,  $Q_{\rm sca}(j)$ were already shown in Fig.~\ref{fig2}. The conclusions are essentially the same as for DLCA clusters (see Sec.~\ref{s41}), viz., single scattering is expected for all the parameter sets, expect for $(x=1.0,~n=3.0)$.

\begin{table}[!htb]
\begin{center}
\begin{tabular}{|c|c|c|c|}
\hline
$x$ \textbackslash ~ $n$ & 1.05 & 1.5 & 3.0 \\ \hline
0.1 &\bf{0.004} & \bf{0.035} & \bf{0.086} \\ \hline
0.5 & \bf{0.028} & \bf{0.252} & 0.622 \\ \hline
1.0 & \bf{0.055} & 0.488 & 1.167 \\ \hline
\end{tabular}
\end{center}
\caption{Analogous to Table~\ref{table1}, but for DLA clusters of $N=384$ particles.}
\label{table2}
\end{table}

\begin{figure}[htb]
\centering
\includegraphics[ width=0.4\linewidth]{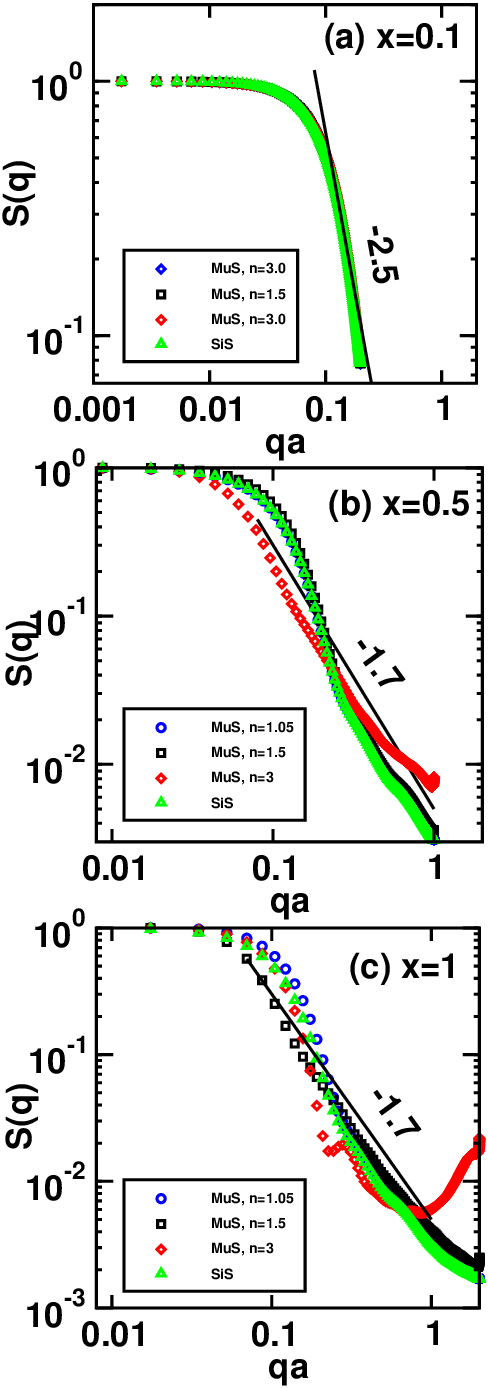}
\caption{Analogous to Fig.~\ref{fig1}, but for DLA clusters. The line of slope $-1.7$ (see text) is drawn on the range $a/R_g < q a < 1$.}
\label{fig3}
\end{figure}

The structure factors for the DLA cluster are shown in Fig.~\ref{fig3}. One can draw the following conclusions, which will be discussed in greater detail in Sec.~\ref{sec5}. \\

1) For $x=0.1$ [Fig.~\ref{fig3}(a)], dipolar single-scattering is the relevant process, in agreement with Table~\ref{table2}. The TMatrix structure factor is numerically indistinguishable from the single-scattering $S(q)$ calculated using Eq.~(\ref{2}) and plotted on the figure as the solid line. Note that the fractal regime is not easily accessible in these numerical experiments, due to the natural limitation in the range of the $q$-values (the maximum value of $qa$ is $2x$, realized in the backward scattering direction). \\
2) For $x=0.5$ [Fig.~\ref{fig3}(b)], dipolar single-scattering is valid for the weak scattering cases with $n=1.05$ (comparison can be done with the solid line or with the corresponding single-scattering DLA structure factor as reproduced in \cite{oh.soren,nisha}). Hence, an excellent agreement between the TMatrix results and the single-scattering data is observed. However, the fractal scaling law in Eq. (\ref{0}), which should appear in the range $0.07 < qa < 1$ is hardly seen. The case $n=1.5$ is also in agreement with the single-scattering result, but through the explanation given in Section~\ref{s31} as the \emph{dipolar} mean-field multiple scattering case holds (see Table~\ref{table2} and Fig.~\ref{fig1}(a)). \\
3) For $x=0.5$, when $qa > 0.2$,  one can note that a power-law behavior:  $S(q) \propto (qa)^{-\delta}$ appears with an anomalous exponent $\delta \simeq 1.7$, substantially different from the fractal dimension $d_f = 2.5$. This behavior was recently discussed in detail by Oh and Sorensen \cite{oh.soren} and  Katyal et al. \cite{nisha}. We conclude that it is a \emph{robust} feature, which is not limited to the dipolar single-scattering case.  Since the scaling behavior is seen in both single-scattering and multiple-scattering conditions, this is a signature of the local arrangement of the particles in the DLA clusters.\\
4) For the parameters $(x=0.5,~n=3.0)$, we expect relevant multipolar mean-field multiple scattering (see Table~\ref{table2} and Fig.~\ref{fig1}(b)). However, the major features of $S(q)$ are similar to the case $(x=0.5,~n=1.5)$, though appearing at systematically smaller values of $q$, i.e., for larger characteristic lengths. In other words, extracting the value of the gyration radius from the Guinier regime [$S(q) \sim \exp(-q^2R_g^2/3)$] would result in an over-estimation by some $50\%$ of its correct value. The anomalous power-law: $S(q) \propto (qa)^{-\delta}$ appears again in this case with the same exponent $\delta \simeq 1.7$ as for the above case.\\
5) For $x=1.0$ [Fig.~\ref{fig3}(c)], the change with increasing refractive index, $n$ is similar to the case $x=0.5$, but occurs at smaller values of $n$. The case $(x=1.0,n=1.05)$ is comparable to the single-scattering result, though it exhibits a shorter  Guinier regime. The case $(x=1.0,~n=1.5)$ emphasizes the appearance of an anomalous power-law for $S(q)$.\\
6) The behavior arising for $(x=1.0,~n=3.0)$ is completely different. Here, the mean-field approximation is no longer valid (see Table~\ref{table2} and Fig.~\ref{fig1}(d)). In that case, one can see strong scattering in both the forward and the backward directions (small and large values of $q$), which destroys the possible scaling law for $S(q)$. 

\section{Scaling Behavior of the Structure Factor: Further Discussion}
\label{sec5}

The TMatrix data analyzed in Section~\ref{s4} shows different scenarios for scattering aggregates with fractal dimension $d_f < 2$ and $d_f > 2$. This is in agreement with a qualitative argument by Berry and Percival \cite{berry} who stated that the electromagnetic scattering features {\it must} differ essentially in these cases. This statement holds true whether the geometric projection of the aggregate on the plane perpendicular to the incident wave direction is compact ($d_f > 2$), or still a fractal ($d_f < 2$).

The case  $d_f < 2$ exhibits rather simple behavior. The scaling relation of the generalized structure factor, $S(q) \sim (qa)^{-d_f}$, is valid in the fractal regime ($a/R_g < qa < 1$) for moderate multiple-scattering [$(N-1) |\chi A | < 0.45$] and moderate multipolar feature ($|n| x < 1.5$). This confirms the common belief regarding the robustness of the fractal scaling law for $d_f < 2$.  However, in the most extreme case that we studied, i.e., $(x=1.0, n=3.0)$, the backscattering signal is strongly enhanced. This is related to coherent backscattering of a random system made of small particles \cite{petrova}, resulting in weak localization of the wave in the backward direction. This phenomenon leads to a  backscattering cone of enhanced intensity, in particular for fractal aggregates \cite{olivi}, as seen in Fig.~\ref{fig1}(c), and to a smaller extent in Fig.~\ref{fig1}(b) ($n=3.0$). As it corresponds to a localization event, coherent backscattering is not part of the mean-field theory developed in \cite{berry,MFBotet}, thus it is not taken into account in the condition (\ref{cond1}).

For $d_f <2$, we conclude that the structure factor can be fully explained with the help of the Guinier regime, the fractal regime, and the coherent backscattering feature. In particular, the fractal dimension can be safely extracted from the relation (\ref{0}), even if multiple-scattering components are not small.

\begin{figure}[htb]
\centering
\includegraphics[ width=0.4\linewidth]{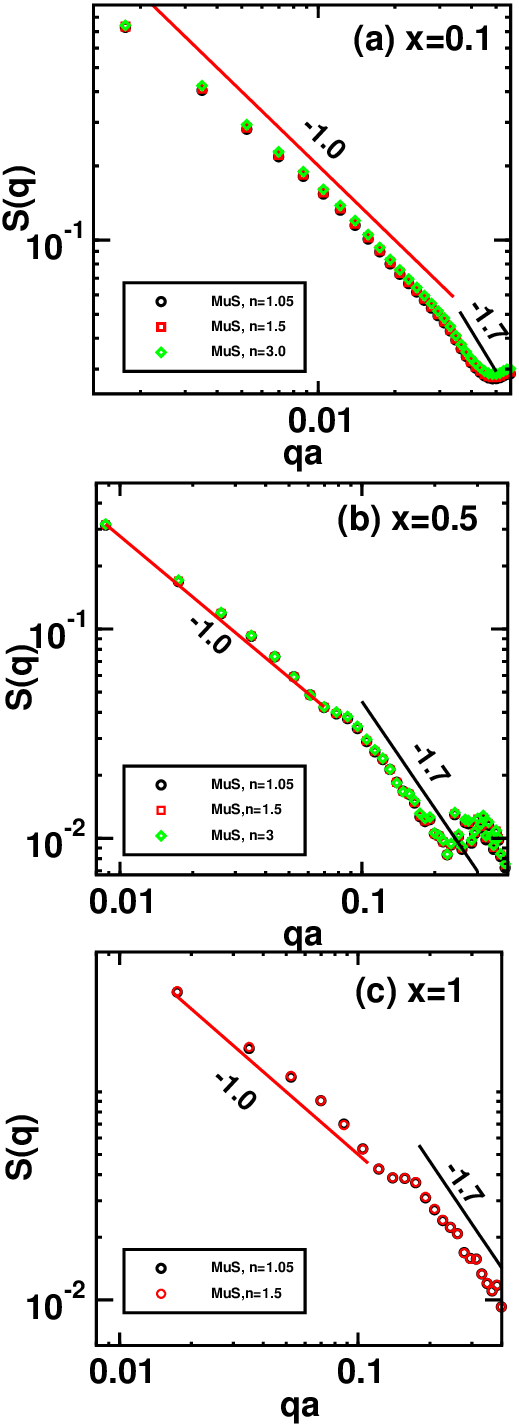}
\caption{Analogous to Fig.~\ref{fig1}, but for linear ($d_f=1$) clusters. The line of slope $-1.7$ (see text) is drawn on the range $a/R_g < q a < 1$.}
\label{fig4}
\end{figure}

We have emphazised the robustness of the power law exponent $-d_f$ in the $S(q)$ vs. $q$ plot for $d_f < 2$. The next question that arises is related to the sensitivity of the fractal domain to a possible enlargement of the backscattering regime. In fact, the backscattering cone angle increases with the mean free length of the wave scattering in the system \cite{petrova}. Thus, the  largest backscattering cones are expected to occur for the smallest fractal dimensions, for which the particles forming the aggregate are far apart. Therefore, we performed TMatrix calculations for a linear chain ($d_f = 1$) of spherical particles with an angular averaging, for the same parameters as used previously. The results are shown in Fig.~\ref{fig4}, and should be compared with the exact result of the single-scattering case, which is $S(q) \sim \pi/(2Nqa)$ for the large aggregates of size $N$ and $Nqa>\sqrt{3}$.

Indeed, multiple scattering is expected to be irrelevant in this case because of the low density of the aggregate, or equivalently the small values of $(N-1) |\chi A |$, which are about 3 times smaller than the values for the DLCA aggregates. However, the results plotted in Fig.~\ref{fig4} show a discrepancy between the single-scattering result (which is a straight line of slope $-1$ on the figures) and the multiple-scattering results - coherent backscattering leads to an enhanced scattering signal upto $qa \sim 0.1$, hiding the fractal regime. Interestingly, the behavior for $qa > 0.1$ can be well represented as an anomalous power-law scaling: $S(q) \sim (qa)^{-\delta}$ with an exponent $\delta \simeq 1.7$. We should emphasize here that this feature is due to the coherent scattering, because it is neither due to special morphology of the aggregate (the anomalous scaling law is not seen in the single-scattering data) nor due to the ordinary multiple-scattering process (because of the small values of $(N-1) |\chi A |$).

We now have the tools to analyze the seemingly more complicated case $d_f > 2$ (shown in Fig.~\ref{fig3}). For small values of the optical size parameter ($x=0.1$), the multiple-scattering process is irrelevant due to the small values of $(N-1) |\chi A |$ (see Table~\ref{table2}). Therefore, the $S(q)$ behavior is the same as for the single-scattering case. The Guinier regime is dominant because of the small size of the aggregate. For $x=0.5$, one should also recover the single-scattering structure factor because of the small values of $(N-1) |\chi A |$. This is indeed the case while comparing Fig.~\ref{fig3}(b) and the corresponding single-scattering figures in \cite{oh.soren,nisha}. The unexpected power-law scaling $S(q) \sim (qa)^{-\delta}$ with the exponent $\delta \simeq 1.7$, as seen for $0.2< qa < 1$, is due to the local morphology of the aggregates, since this is a feature of the single-scattering pattern (thus of the Fourier transform of the correlation function $g(r)$). Because of the slope $-1.7$ being similar to the DLCA fractal slope, it is conjectured that a DLCA-like local arrangement of particles is present inside the DLA clusters \cite{oh.soren}. This can be checked directly for the small-$r$ behavior of the correlation function $g(r)$ \cite{unpub}. 

The present scattering data also strengthens the above conclusion, though the precise value of the slope may be difficult to estimate. For example, the same exponent $\delta$ has been estimated to be $1.2$ (instead of $1.7$) for composite clusters made of DLCA ($d_f=1.8$) at short length-scales \cite{riviere}, and DLA ($d_f=2.5$) at long length-scales (with parameters $x=0.354$, $n=1.4607$, $N=4275$ and using the DDA code \cite{draine}). The latter case is comparable to our $(x=0.5, n=1.5)$ case.  As explained in Ref.~\cite{oh.soren}, the hump seen for $qR_g \simeq 1$ is due to a misfit between the Guinier regime  for the small values of $q$ [$S(q) \sim \exp  (-4 (qa)^2 N^{2/2.5}/3)$] which uses the fractal dimension $2.5$ to obtain the overall gyration radius. Also, the DLCA-like fractal regime [$S(q) \sim 1/(N(qa)^{1.78})$] uses the fractal dimension $1.78$ for the large $q$. This hump in the structure factor results in a kind of {\it corona} as appearing for a porous sphere of fractal dimension $d_f=3$ with a high refractive index \cite{laven}. The general appearance of the hump for the DLA clusters provides another proof that the Guiner and fractal regimes are robust against moderate multiple-scattering processes.

To summarize, one can note two special behaviors for larger values of $x|n|$, as described below. \\
1) For $(x=0.5,~n=3)$, our data is consistent with the fractal dimension $-2.5$, followed by an anomalous scaling law (slope $-1.7$). The same behavior occurs for the case $(x=1,~n=1.5)$ (which yields the same value of $x|n|$). The cause of this effect is probably due to the multiple scattering process as evident from the relatively large value of $(N-1) |\chi A | \simeq 0.6$.  However, further investigation is required to understand its origin and behavior. In this particular example, we can also see the onset of coherent backscattering. \\
2) For the $x=1$ case, the most striking feature is the coherent backscattering enhancement developing for the largest values of the refractive index, for both the DLA [Fig.~\ref{fig1}(c)] and DLCA [Fig.~\ref{fig3}(c)] cases.

\section{Conclusion}
\label{s6}

We have analyzed TMatrix scattering results for three different classes of fractal aggregates (the linear chain, the DLCA and DLA clusters) for various optical size parameters and refractive index in the domain of common applications. We conclude that the fractal regime, allowing for direct estimation of the fractal dimension, is robust against moderate multiple scattering. However, its range (in the $q$-values) can be significantly reduced either by a specific local arrangement of the particles inside the scatterer (DLA case) or by a coherent backscattering process (for small fractal dimensions).

We provide an explicit criterion to check whether a scattering experiment is in the single or multiple-scattering regime. We emphasize that the criterion is valid only in the \emph{fractal regime}, and not in the forward or backward scattering directions. Thus, we now have a tool to extract the fractal dimension in a controlled way from the scattering data in the fractal regime.
\\

\subsection*{Acknowledgments} 

R.B. acknowledges the France-India CEFIPRA project No 4407-A ``Cometary grains: observations and simulations'' which enabled the onset of this work. N.K. acknowledges a UGC-India fellowship.





\end{document}